\documentclass[11pt]{article}

\usepackage{amsmath}
\usepackage{amsfonts}
\usepackage{graphicx}
\usepackage{epsfig}

\newtheorem{proposition}{Proposition}
\newtheorem{lemma}{Lemma}
\newtheorem{remark}{Remark}
\newtheorem{exn}{Example}

\def\5n{\negthinspace \negthinspace \negthinspace \negthinspace \negthinspace }
\def\4n{\negthinspace \negthinspace \negthinspace \negthinspace }
\def\3n{\negthinspace \negthinspace \negthinspace }
\def\2n{\negthinspace \negthinspace }
\def\n{\negthinspace }
\def\ds{\displaystyle}
\def\proof {{Proof.} }
\def\endproof{\hfill $\Box$ \vskip .3cm}
\def\eqbydef{\mathrel{\stackrel{\Delta}{=}}}

\def\e{\mathbf{e}}

\def\u{\mathbf{u}}
\def\P{\mathbf{P}}
\def\0{\mathbf{0}}
\def\1{\mathbf{1}}
\def\E{\mathbb{E}}
\def\K{\mathbf{K}}

\begin{document}

\title{Unified Framework of Mean-Field Formulations for Optimal Multi-period Mean-Variance Portfolio Selection\thanks{This work was partially supported by Research Grants Council of Hong Kong under grants 414207 and 520412, and by National Natural Science Foundation of China under grant 71201094.}}

\author{Xiangyu Cui\thanks{School of Statistics and Management, Shanghai University of Finance and Economics, Shanghai, China. E-mail: cui.xiangyu@mail.shufe.edu.cn.}, \quad
Xun Li\thanks{Department of Applied Mathematics, The Hong Kong Polytechnic University, Hong Kong, China. E-mail: malixun@polyu.edu.hk.}  \quad and \quad Duan Li\thanks{Corresponding author. Department of Systems Engineering and Engineering Management, The Chinese University of Hong Kong, Hong Kong, China. E-mail: dli@se.cuhk.edu.hk.}}

\maketitle

\begin{abstract}
The classical dynamic programming-based optimal stochastic control methods
fail to cope with nonseparable dynamic optimization problems as
the principle of optimality no longer applies in such situations.
Among these notorious nonseparable problems, the dynamic
mean-variance portfolio selection formulation had posted a great
challenge to our research community until recently. A few solution
methods, including the embedding scheme, have been developed in
the last decade to solve the dynamic mean-variance portfolio
selection formulation successfully. We propose in this paper a
novel mean-field framework that offers a more efficient modeling
tool and a more accurate solution scheme in tackling directly the issue of nonseparability and deriving the optimal policies
analytically for the multi-period mean-variance-type portfolio
selection problems.

\end{abstract}

\section{Introduction}

The mean-field type of optimal stochastic control models deals
with problems in which both the system dynamics and objective
functional could involve the states as well as the \textit{expected values
of the states}. The past few years have witnessed an increasing number of successful
applications of the mean-field formulation, including mean-field type of stochastic control problems,
 in various fields of science, engineering, financial management, and economics.
Although the research in this direction
has been well developed for continuous-time control problems, it
lacks progress in both theoretical investigation and applications
in discrete-time problems. The current work in this paper aims to
employ the mean-field formulation to cope with seemingly
non-tractable nonseparability in discrete-time portfolio selection
problems. In particular, we revisit three challenging, yet
practically important, portfolio selection models over a
finite-time investment horizon (see Li and Ng \cite{LiNg2000},
Costa and Nabholz \cite{CostaNabholz}, Zhu et al.
\cite{Zhu:2004}), reformulate them as discrete-time
linear-quadratic control problems of a mean-field type, and derive
their optimal strategies with improved solution qualities.

Since Markowitz \cite{Markowitz} published his seminal work on the
mean-variance portfolio selection sixty years ago, the mean-risk
portfolio selection framework has become one of the most
significant ingredients in the modern financial theory.  An
important yet essential research theme under the mean-risk
portfolio selection framework is to strike a balance between
achieving a high mean of the investment return and minimizing the
corresponding risk. If we adopt the variance of the terminal
wealth as a risk measure for investment, we have the following
mathematical formulations of the classical \textit{static}
mean-variance models,
\begin{align*}
(MV(\sigma)) \quad \max & ~~\mathbb{E}(x_1),  \\
\mbox{s.t.} &~ \mbox{\rm Var}(x_1)\leq \sigma^2,\\
&~x_1=x_0+u_0\bullet S_0,
\end{align*}
and
\begin{align*}
(MV(\epsilon)) \quad \min &~~\mbox{\rm Var}(x_1), \\
\mbox{s.t.} &~\mathbb{E}(x_1)\geq \epsilon,\\
&~ x_1=x_0+u_0\bullet S_0,
\end{align*}
which are equivalent to
\begin{align*}
(MV_s)\quad \max &~~\mathbb{E}(x_1)-\omega\mbox{\rm Var}(x_1),\\
\mbox{s.t.} &~~x_1=x_0+u_0\bullet S_0,
\end{align*}
where $x_t$ is the wealth at time $t$, $u_t$ is the portfolio
strategy at time $t$, $S_t$ is the random return at time $t$,
$x_0+u_0\bullet S_0$ denotes the random terminal wealth $x_1$ from
applying strategy $u_0$ in the market with initial wealth $x_0$,
and $\omega>0$ denotes the trade-off between the two conflicting
objectives of maximizing expected return and minimizing the risk.
The optimal portfolio strategy and solution scheme of $(MV_s)$ can
be found in Merton \cite{Merton} when shorting is allowed and in
Markowitz \cite{Markowitz} when shorting is prohibited.

However, the extension to a dynamic version of mean-variance portfolio selection was blocked for four decades until recently. Let us consider the following abstract form for the dynamic mean-variance portfolio selection problem,
\begin{align*}
(MV(\omega))\quad \max_u &~~\mathbb{E}(x_T)-\omega\mbox{\rm Var}(x_T), \\
\mbox{s.t.} & ~~x_T=x_0+\{u_t\bullet S_t\}\mid_{t=0}^{T-1},
\end{align*}
where $x_0+ \{u_t\bullet S_t\}\mid_{t=0}^{T-1}$ denotes the random
terminal wealth $x_T$ from applying strategy $\{u_t\}
\mid_{t=1}^{T-1}$ in the market with initial wealth $x_0$. Due to
the non-smoothing property of the variance term, i.e.,
\begin{align*}
\mbox{\rm Var}\big(\mbox{\rm Var}(\cdot|\mathcal{F}_i)|\mathcal{F}_j\big)\neq \mbox{\rm Var}(\cdot|\mathcal{F}_j),\quad \forall i<j,
\end{align*}
where $\mathcal{F}_j$ is the information set available at time $j$
and $\mathcal{F}_{j-1}\subset \mathcal{F}_j$, $(MV(\omega))$ is
not a standard stochastic control problem whose objective functional
involves the wealth state as well as a nonlinear function of the
expected wealth and, thus, does not satisfy the principle of
optimality. Therefore, all the traditional dynamic programming-based
 optimal stochastic control solution methods  no longer apply.

We now briefly summarize the main approaches in the current literature to overcome the
difficulty resulted from the nonseparability.
Adopting an embedding scheme, Li and Ng \cite{LiNg2000} and Zhou and
Li \cite{ZhouLi2000} considered the following family of auxiliary
problems, $\mathcal{A}(\omega,\lambda)$, parameterized in parameter
$\lambda$,
\begin{align*}
\mathcal{A}(\omega,\lambda)\quad \min_u & ~~\mathbb{E}(\omega x_T^2- \lambda x_T), \\
\mbox{s.t.} & ~~x_T=x_0+\{u_t\bullet S_t\}\mid_{t=0}^{T-1}.
\end{align*}
Note that problem $\mathcal{A}(\omega,\lambda)$ is a separable
linear-quadratic stochastic control (LQSC) formulation and can be
thus solved analytically. Li and
Ng \cite{LiNg2000} and Zhou and Li \cite{ZhouLi2000} derived the
optimal policy to the primal nonseparable problem $(MV(\omega))$
via identifying the optimal parameter $\lambda$ and applying the
optimal $\lambda^*$ to $\mathcal{A}(\omega,\lambda)$. The embedding
scheme has been also extended to multi-period mean-variance model
with intertemporal restrictions (see Costa and Nabholz
\cite{CostaNabholz}), multi-period mean-variance model in a
stochastic market whose evolution is governed by a Markovian chain
(see \c{C}elikyurt and \"{O}zekici \cite{Celikyurt-Ozekici}), a
generalized mean-variance model with risk control over bankruptcy
(see Zhu et al. \cite{Zhu:2004}), and dynamic mean-variance
asset-liability management (see Leippold et al. \cite{Leippold},
Chiu and Li \cite{ChiuLi}, Chen and Yang \cite{ChenYang}).

By introducing an auxiliary variable $d$ and an equality
constraint $\mathbb{E}(x_T)=d$ for the expected terminal wealth,
Li et al. \cite{LZL} paved the road to study the following
slightly modified, albeit equivalent, version of $(MV(\omega))$
(we omit the no-shorting constraint here and focus on the model
itself),
\begin{align*}
(MV(d))\quad \min_u &~~ \mbox{\rm Var}(x_T)=\mathbb{E}(x_T-d)^2, \\
\nonumber \mbox{s.t.} &~~ \mathbb{E}(x_T)=d, \\
\nonumber & ~~ x_T=x_0+\{u_t\bullet S_t\}\mid_{t=0}^{T-1}.
\end{align*}
Introducing a Lagrangian multiplier $\lambda$ and applying
Lagrangian relaxation to $(MV(d))$ gives rise to the following
LQSC problem,
\begin{align}
(L(\lambda))\quad \min &~~ \mathbb{E}(x_T-d)^2-\lambda \mathbb{E}(x_T-d), \\
\nonumber \mbox{s.t.} & ~~ x_T=x_0+\{u_t\bullet
S_t\}\mid_{t=0}^{T-1}.
\end{align}
The optimal policy of $(MV(d))$ can be
obtained by maximizing the dual function $L(\lambda)$ over all
Lagrangian multiplier $\lambda\in \mathbb{R}$. In fact, the
Lagrangian problem $(L(\lambda))$ can be further written as the following LQSC
problem,
\begin{align}
(MVH(m))\quad \min &~~ \mathbb{E}(x_T-m)^2, \\
\nonumber \mbox{s.t.} & ~~x_T=x_0+\{u_t\bullet
S_t\}\mid_{t=0}^{T-1},
\end{align}
where $m=d+\lambda/2$. Problem $(MVH(m))$ is a special
mean-variance hedging problem, in which an investor hedges the
target $m$ by his/her portfolio under a quadratic objective
function. Problem $(MVH(m))$ has been well studied and can be
solved by LQSC theory (see Li et al. \cite{LZL}),
martingale/convex duality theory (see Schweizer
\cite{Schweizer1996}, Xia and Yan \cite{XiaYan2006}) and
sequential regression method (see \v{C}ern\'{y} and Kellsen
\cite{CernyKallsen2009}).

In all the literature mentioned above, a static optimization
procedure is always necessary to identify an optimal parameter in
the parameterized auxiliary problem $\mathcal{A}(\omega,\lambda)$,
$(L(\lambda))$ or $(MVH(m))$. Actually, based on the pure geometric
structure of $(MV(\omega))$, Sun and Wang \cite{SunWang} proved that
the optimal terminal wealth $x_T^*$ takes the following form,
\begin{align*}
x_T^*=x_0+\frac{1}{2\omega}\frac{1}{\mathbb{E}(1-\{\varphi^*_t\bullet S_t\}\mid_{t=0}^{T-1})}\{\varphi^*_t\bullet S_t\}\mid_{t=0}^{T-1},
\end{align*}
where $\varphi^*$ is the policy of the following particular
mean-variance hedging problem,
\begin{align*}
(MVH(1)) ~~\min &~~ \mathbb{E}(x_T-1)^2, \\
\mbox{s.t.} & ~~x_T= \{\varphi_t\bullet S_t\}\mid_{t=0}^{T-1}.
\end{align*}

All the above approaches attempt to embed the ``nontractable''
nonseparable mean-variance portfolio selection problem into a
family of tractable LQSC problems. Although these transformations
seem necessary, one meaningful yet challenging question emerges
naturally: Are we able to \textit{directly} tackle the above
nonseparable dynamic mean-variance problems (without introducing
an auxiliary problem)?

The mean-variance problem is in fact a special case of the
mean-field type problems where both the underlying dynamic system and
the objective functional involve state processes as well as their
expected values (hence the name mean-field). This critical feature
differentiates the mean-variance problem from standard stochastic
control problems. The theory of the \textit{mean-field stochastic
differential equation} can be traced back to Kac \cite{Kac} who
presented the McKean-Vlasov stochastic differential equation
motivated by a stochastic toy model for the Vlasov kinetic equation
of plasma. Since then, the research on related topics and their
applications has become a notable and serious endeavor among
researchers in applied probability and optimal stochastic controls,
particularly in financial engineering. This new direction, however,
requires new analytical tools and solution techniques. For instance,
in a recent research on mean-field forward stochastic LQ optimal
control problems, Yong \cite{Yong2011} introduced a system of two Riccati equations to solve the problem.
Representative works in mean-field include, but not limited to,
Mckean \cite{McKean}, Dawson \cite{Dawson}, Chan \cite{Chan},
Buckdahn et al. \cite{Buckdahn-Li-Peng}, Borkar and Kumar
\cite{Borkar-Kumar}, Crisan and Xiong \cite{Crisan-Xiong}, Andersson
and Djehiche \cite{Andersson-Djehiche}, Buckdahn et al. \cite{Buckdahn-Djehiche-Li},
Meyer-Brandis et al. \cite{Meyer-Brandis-Oksendal-Zhou},
Nourian et al. \cite{Nourian-Caines-Malhame-Huang} and Yong \cite{Yong2011}.
Despite active research efforts on mean-field in recent years, the
topic of multi-period models in discrete-time remains a relatively
unexplored subject where the mean-field modeling scheme has not yet
been applied.

In this paper, we will develop a unified framework of mean-field
formulations to investigate three multi-period mean-variance models in
the literature: classical multi-period mean-variance model in Li
and Ng \cite{LiNg2000}, multi-period mean-variance model with
intertemporal restrictions in Costa and Nabholz
\cite{CostaNabholz}, and a generalized mean-variance model with
risk control over bankruptcy in Zhu et al. \cite{Zhu:2004}. We
demonstrate that the mean-field approach represents a new promising
way in dealing with nonseparable stochastic control problems related
to the mean-variance formulations and even improves solution
quality of some existing results in the literature.

\section{Mean-Field Formulations for Multi-Period Mean-Variance Portfolio Selection}

We consider in this paper a capital market consisting of one
riskless asset and $n$ risky assets within a time horizon $T$. Let
$s_t ~(>1)$ be a given deterministic return of the riskless asset
at period $t$ and $\e_t = [e_t^1,\cdots,e_t^n]'$ the vector of
random returns of the $n$ risky assets at period $t$. We assume
that vectors $\e_t$, $t$ = 0, 1, $\cdots$, $T$ $-$ 1, are
statistically independent and the only information known about the
random return vector $\e_t$ is its first two moments, its mean
$\mathbb{E}(\e_t)=[\mathbb{E}(e_t^1),  \mathbb{E}(e_t^2), \cdots,
\mathbb{E}(e_t^n)]^{\prime }$ and its positive definite covariance
\[
\mbox{\rm Cov}\left( \e_t\right) =\mathbb{E}(\e_t\e_t')-\mathbb{E}(\e_t)\mathbb{E}(\e_t')=\left[
\begin{array}{ccc}
\sigma _{t,11}  & \cdots & \sigma _{t,1n} \\
\vdots & \ddots & \vdots \\
\sigma _{t,1n} & \cdots & \sigma _{t,nn}
\end{array}
\right]\succ 0.
\]
From the above assumptions, we have
\begin{align}\nonumber
 \left[\begin{array}{cc} s_t^2 & s_t\mathbb{E}(\e_t') \\
s_t \mathbb{E}(\e_t) & \mathbb{E}(\e_t\e_t')
 \end{array}\right]
 \succ 0.
\end{align}

We further define the excess return vector of risky assets $\P_t$ as
\[
\P_t=[P^1_t,P^2_t,\cdots,P^n_t]'=[(e^1_t-s_t),(e^2_t-s_t),\cdots,(e^n_t-s_t)]'.
\]
The following is then true for $t=0,1,\dots,T-1$:
\begin{align*}
\left[\begin{array}{cc} s_t^2 & s_t\mathbb{E}(\P_t') \\
s_t \mathbb{E}(\P_t) & \mathbb{E}(\P_t\P_t')
 \end{array}\right] =\left[\begin{array}{cccc} 1 & \0' \\
 -\1 & I\\
 \end{array}\right]
 \left[\begin{array}{cc} s_t^2 & s_t \mathbb{E}(\mathbf{e_t}') \\
s_t \mathbb{E}(\mathbf{e_t}) & \mathbb{E}(\mathbf{e_te_t}')
 \end{array}\right]
\left[\begin{array}{cccc} 1 & \0' \\
 -\1 & I\\
 \end{array}\right]\succ 0,
\end{align*}
where $\1$ and $\0$ are the $n$-dimensional all-one and all-zero vectors, respectively, and $I$ is the $n \times n$ identity matrix, which further implies
\begin{align*}
\mathbb{E}(\P_t\P_t')\succ 0, \quad \forall t=0,1,\cdots,T-1,\\
s_t^2(1-B_t)>
0, \quad \forall t=0,1,\cdots,T-1,
\end{align*}
where $B_t\eqbydef\mathbb{E}(\P_t')\mathbb{E}^{-1}(\P_t\P_t')\mathbb{E}(\P_t)$.

An investor joins the market at the beginning of period $0$ with
an initial wealth $x_0$. He/she allocates $x_0$ among the riskless
asset and $n$ risky assets at the beginning of period $0$ and
reallocates his/her wealth at the beginning of each of the
following $\left( T-1\right)$ consecutive periods. Let $x_t$ be
the wealth of the investor at the beginning of period $t$, and
$u_t^i$, $i = 1, 2, \cdots, n$, be the amount invested in the
$i$-th risky asset at period $t$. Then, $x_t-\sum_{i=1}^n u_t^i$
will be the amount invested in the riskless asset at period $t$.
The information set at the beginning of period $t$ is denoted as
$$\mathcal{F}_t=\sigma(\mathcal{F}_0\vee\sigma(\mathbf{P}_0,\mathbf{P}_1,\cdots,\mathbf{P}_{t-1})),$$
where $\mathcal{F}_0$ contains $x_0$, $s_t$ and the first and
second moment information of $\mathbf{P}_t$, $t = 0, 1, \cdots,
T-1$. We confine an admissible investment strategies to be
$\mathcal{F}_t$-measurable Markov control, i.e., $\u_t\in\mathcal{F}_t$. Then,
$\P_t$ and $\u_t$ are independent, $\{x_t\}$ is an adapted
Markovian process and
$\mathcal{F}_t=\sigma(\mathcal{F}_0\vee\sigma(x_t))$.


The conventional multi-period mean-variance model is to seek the best
strategy, $\mathbf{u}_t^*=[(u_t^1)^*,(u_t^2)^*, \cdots$,
$(u_t^n)^*]'$, $t = 0, 1, \cdots, T - 1$, which is the optimizer
of the following stochastic discrete-time optimal control problem,
\begin{align}
(MMV)~~\max &~~ \mathbb{E}(x_T) - \omega_T\mbox{\rm Var}(x_T), \\
\nonumber {\rm s.t.} &~~ x_{t+1} = \sum_{i=1}^ne_t^iu_t^i+\bigg( x_t-\sum_{i=1}^nu_t^i\bigg) s_t \\
\label{wealth} &~~ \quad \quad =s_tx_t+\P_t'\mathbf{u}_t, \quad t=0,1,\cdots,T-1,
\end{align}
where $\omega_T>0$ is the trade-off parameter between the mean and the variance of the terminal wealth.

The multi-period mean-variance model with intertemporal restrictions is
to find the optimal control of the following problem,
\begin{align*}
(MMV-IR)~~\max &~~ \sum_{t\in I_\alpha}\alpha_t\left[\ell_t\mathbb{E}(x_t) - \rho_t\mbox{\rm Var}(x_t)\right], \\
{\rm s.t.} &~~ x_{t+1} = s_tx_t+\P_t'\mathbf{u}_t, \quad t=0,1,\cdots,T-1,
\end{align*}
where $I_{\alpha}=\{\tau_1,\cdots,\tau_\alpha\}$ with
$\tau_\alpha=T$ is the set of time instances on which the investor
evaluates the performance of the portfolio, $\alpha_t\ell_t$ and
$\alpha_{t}\rho_t>0$ are the time-$t$ weights of the mean and the
variance in the objective functional. In particular, if we choose
$I_{\alpha}=\{T\}$, $\alpha_T\ell_T=1$ and
$\alpha_T\rho_T=\omega_T>0$, $(MMV-IR)$ reduces to the
conventional multi-period mean-variance portfolio selection model
$(MMV)$ studied in Li and Ng \cite{LiNg2000}. If $I_{\alpha}$
contains time instances other than $T$, $(MMV-IR)$ is the
multi-period portfolio selection problem with intertemporal
restrictions considered in Costa and Nabholz \cite{CostaNabholz}.
Without loss of generality, we let $I_{\alpha}$ include all time
instants from 0 to $T$, while setting some
$\alpha_t=\ell_t=\rho_t=0$ for these time instances which do
not need to be evaluated.

The generalized mean-variance model for dynamic portfolio selection with
risk control over bankruptcy is formulated as
\begin{align*}
(MMV-B)~~\max &~~ \E(x_T)-\omega_T{\rm Var}(x_T), \\
\mbox{s.t.} &~~x_{t+1} = s_tx_t+\P_t'\mathbf{u}_t, \quad t=0,1,\cdots,T-1, \\
&~~P(x_t\leq b_t) \leq a_t,\quad t=1,2,\cdots,T-1,
\end{align*}
where $b_t$ is the disaster level and $a_t$ is the acceptable maximum
probability of bankruptcy set by the investor. By Tchebycheff
inequality, problem $(MMV-B)$ can be transformed into the following
$(GMV)$ model (see Zhu et al. \cite{Zhu:2004}),
\begin{align*}
(GMV)~~\max &~~ \E(x_T)-\omega_T{\rm Var}(x_T), \\
\mbox{s.t.} &~~x_{t+1} = s_tx_t+\P_t'\mathbf{u}_t, \quad t=0,1,\cdots,T-1, \\
&~~{\rm Var}(x_t)\leq a_t\left[\E(x_t)-b_t\right]^2,\quad t=1,2,\cdots,T-1.
\end{align*}
To solve $(GMV)$, let us consider the Lagrangian maximization problem,
\begin{align*}
(L(\omega))~~\max &~~ \E(x_T)-\omega_T{\rm Var}(x_T)-\sum_{t=1}^{T-1}\omega_t\left[{\rm Var}(x_t)-a_t\left(\E(x_t)-b_t\right)^2\right], \\
\mbox{s.t.} &~~x_{t+1} = s_tx_t+\P_t'\mathbf{u}_t, \quad t=0,1,\cdots,T-1,
\end{align*}
where $\omega=(\omega_1,\omega_2,\cdots,\omega_{T-1})^\prime\in\mathbb{R}_{+}^{T-1}$ is the vector of Lagrangian multipliers.

We are now building up the mean-field formulations for problems
$(MMV-IR)$ and $(L(\omega))$, respectively. For
$t=0,1,\cdots,T-1$, the evolution of the expectation of the wealth
dynamics specified in (\ref{wealth}) can be presented as
\begin{align}\label{eqn_expect_wealth}
\left\{\begin{array}{rcl}
\mathbb{E}(x_{t+1}) & \3n=\3n & s_t\mathbb{E}(x_t)+\mathbb{E}(\P_t')\mathbb{E}(\u_t), \\
\mathbb{E}(x_0)     & \3n=\3n & x_0,
\end{array}\right.
\end{align}
due to the independence between $\P_t$ and $\u_t$. Combining
(\ref{wealth}) and (\ref{eqn_expect_wealth}) yields the following
for $t=0,1,\cdots,T-1,$
\begin{align}\label{eqn_wealth_minus_expect}
\left\{\begin{array}{rcl}
x_{t+1}-\mathbb{E}(x_{t+1}) & \3n=\3n & s_t\big(x_t-\mathbb{E}(x_t)\big) + \P_t'\u_t - \mathbb{E}(\P_t')\mathbb{E}(\u_t) \\
                            & \3n=\3n & s_t\big(x_t-\mathbb{E}(x_t)\big) + \P_t'\big(\u_t-\mathbb{E}(\u_t)\big) + \big(\P_t'-\mathbb{E}(\P_t')\big)\mathbb{E}(\u_t), \\
x_0 - \mathbb{E}(x_0)       & \3n=\3n & 0.
\end{array}\right.
\end{align}

What we are actually doing here is to enlarge the state space
$(x_t)$ into $(\E(x_t), x_t-\E(x_t))$ and the control space
$(\u_t)$ into $(\E(\u_t), \u_t-\E(\u_t))$. Although control vector
$\E(\u_t)$ and $\u_t-\E(\u_t)$ can be decided independently at
time $t$, they should be chosen such that
\begin{align*}
\E(\u_t-\E(\u_t))=\0,\quad t=0,1,\cdots,T-1.
\end{align*}
We also confine admissible investment strategies $(\E(\u_t),
\u_t-\E(\u_t))$ to be $\mathcal{F}_t$-measurable Markov control. Then,
$\{(\E(x_t), x_t-\E(x_t))\}$ is again an adapted Markovian process
and $\mathcal{F}_t=\sigma(\mathcal{F}_0\vee\sigma(\E(x_t),
x_t-\E(x_t)))$.

The problem $(MMV-IR)$ can be reformulated as a mean-filed type of linear quadratic optimal stochastic control problem,
\begin{align*}
(MMV-MF)~~\max &~~ \sum_{t=1}^{T}\alpha_t\Big\{\ell_t\mathbb{E}(x_t)-\rho_t\mathbb{E}\big[(x_t - \mathbb{E}(x_t))^2\big]\Big\}, \\
\mbox{s.t.} &~~\mathbb{E}(x_{t}) \mbox{ satisfies dynamic equation (\ref{eqn_expect_wealth})}, \\
&~~x_{t+1}-\mathbb{E}(x_{t}) \mbox{ satisfies dynamic equation (\ref{eqn_wealth_minus_expect})}, \\
&~~\E(\u_t-\E(\u_t))=\0,\quad t=0,1,\cdots,T-1.
\end{align*}
Similarly, problem $(L(\omega))$ can be reexpressed as
\begin{align*}
(L-MF(\omega))~~\max &~~ \E(x_T)-\omega_T\E\big[(x_T - \mathbb{E}(x_T))^2\big] \\
&~~-\sum_{t=1}^{T-1}\omega_t\Big\{\E\big[(x_t - \mathbb{E}(x_t))^2\big]-a_t(\E(x_t)-b_t)^2\Big\}, \\
\mbox{s.t.} &~~\mathbb{E}(x_{t}) \mbox{ satisfies dynamic equation (\ref{eqn_expect_wealth})}, \\
&~~x_{t+1}-\mathbb{E}(x_{t}) \mbox{ satisfies dynamic equation (\ref{eqn_wealth_minus_expect})},\\
&~~\E(\u_t-\E(\u_t))=\0,\quad t=0,1,\cdots,T-1.
\end{align*}
In the above two formulations of a mean-field type, the
corresponding problems become separable linear quadratic optimal
stochastic control problems in the expanded state space $(\E(x_t),
x_t-\E(x_t))$ with the second control vector $\u_t-\E(\u_t)$ being
constrained by a linear equation.

\section{Optimal Policies for Multi-period Mean-Variance Portfolio Selection with and without Intertemporal Restrictions}

\begin{lemma}[Sherman-Morrison formula]\label{lem_S_M}
\sl
Suppose that $A$ is an invertible square matrix and $\mu$ and $\nu$ are two given vectors. If
\begin{align*}
1+\nu'A^{-1}\mu\neq 0,
\end{align*}
then the following holds,
\begin{align*}
(A+\mu\nu')^{-1}=A^{-1}-\frac{A^{-1}\mu\nu'A^{-1}}{1+\nu'A^{-1}\mu}.
\end{align*}
\end{lemma}

\begin{lemma}\label{lem:EP}
\sl
Let
$B_t=\mathbb{E}(\P_t')\mathbb{E}^{-1}(\P_t\P_t')\mathbb{E}(\P_t)$.
Then
\begin{align*}
\big[\mathbb{E}(\P_t\P_t')-\mathbb{E}(\P_t)\mathbb{E}(\P_t')\big]^{-1}\mathbb{E}(\P_t)=\frac{\mathbb{E}^{-1}(\P_t\P_t')\mathbb{E}(\P_t)}{1-B_t}.
\end{align*}
\end{lemma}

\proof Applying Sherman-Morrison formula gives rise to the following,
\begin{align*}
&~\big[\mathbb{E}(\P_t\P_t')-\mathbb{E}(\P_t)\mathbb{E}(\P_t')\big]^{-1}\mathbb{E}(\P_t)\\
=&~\left[\mathbb{E}^{-1}(\P_t\P_t')+\frac{\mathbb{E}^{-1}(\P_t\P_t')\mathbb{E}(\P_t)\mathbb{E}(\P_t')\mathbb{E}^{-1}(\P_t\P_t')}{1-\mathbb{E}(\P_t')\mathbb{E}^{-1}(\P_t\P_t')\mathbb{E}(\P_t)}\right]\mathbb{E}(\P_t)\\
=&~\left[\mathbb{E}^{-1}(\P_t\P_t')+\frac{\mathbb{E}^{-1}(\P_t\P_t')\mathbb{E}(\P_t)\mathbb{E}(\P_t')\mathbb{E}^{-1}(\P_t\P_t')}{1-B_t}\right]\mathbb{E}(\P_t)\\
=&~\frac{\mathbb{E}^{-1}(\P_t\P_t')\mathbb{E}(\P_t)}{1-B_t}.
\end{align*}
\endproof

Consider the following separable multi-period control problem,
\begin{align*}
\max &~~ \E\left[\sum_{t=0}^{T-1}h_t(x_t,v_t) + h_T(x_T)\right], \\
\mbox{s.t.} &~~x_{t+1}=f(x_t,v_t),\quad t=0,1,\cdots,T-1,
\end{align*}
where $x_t$ denotes the state, $v_t$ denotes the control,
$f(x_t,v_t)$ represents the dynamics of the state and $h_t(x_t,
v_t)$ is concave in $v_t$. Based on the principle of optimality in
dynamic programming, the optimal control at time $t$ is derived
from the following recursion of dynamic programming,
\begin{align*}
v_t^*=\arg\n\max_{\5n\5n v_t}\big\{\E[J_{t+1}(x_{t+1};v_0,v_1,\cdots,v_{t})|\mathcal{F}_t]+h_t(x_t,v_t)\big\},
\end{align*}
where $\mathcal{F}_t$ is the information set at time $t$, $(v_0,v_1,\cdots,v_{t})$ is the control sequence before time $t+1$ and
\begin{align*}
J_{t+1}(x_{t+1};v_0,v_1,\cdots,v_{t})=\max_{v_{t+1},\cdots,v_{T-1}}\3n\E\left[\sum_{j=t+1}^{T-1}h_j(x_j,v_{j}) + h_T(x_T)\bigg|\mathcal{F}_{t+1}\right]
\end{align*}
is the benefit-to-go function at time $t+1$.

\begin{lemma}\label{lem:DP}
\sl
Assume that
\begin{align*}
\E[J_{t+1}(x_{t+1};v_0,v_1,\cdots,v_{t})|\mathcal{F}_t]=G_t^1(x_t;v_0,v_1,\cdots,v_{t})+G_t^2(x_t;v_0,v_1,\cdots,v_{t}),
\end{align*}
where $\E[G_t^2(x_t;v_0,v_1,\cdots,v_{t})|\mathcal{F}_0]=0$ holds for any admissible $(v_0,v_1,\cdots,v_{t})$. Then
\begin{align*}
&v_{t}^*=\arg\n\max_{\3n\3n v_t}~\left\{G_t^1(x_t;v_0,v_1,\cdots,v_{t})+h_t(x_t,v_t)\right\}, \\
&J_0(x_0)=\max_{v_0,\cdots,v_{t}}\bigg\{\E[G_t^1(x_t;v_0,v_1,\cdots,v_{t})|\mathcal{F}_0]+\sum_{j=0}^{t} \E[h_j(x_j,v_j)|\mathcal{F}_{0}]\bigg\}, \\
& \quad\quad\quad\quad\quad\quad\quad\quad\quad\quad\quad\quad\quad\quad\quad\quad\quad\quad\quad\quad t = 0,1,\cdots,T-1,
\end{align*}
i.e., $G_t^1(x_t;v_0,v_1,\cdots,v_{t}^*)+h_{t}(x_{t},v_{t}^*)$ can be regarded as the benefit-to-go function at time $t$.
\end{lemma}

\proof Based on the principle of optimality of dynamic
programming, the optimal control sequence on or before time $t+1$
is determined by
\begin{align*}
(v_0^*,v_1^*,\cdots,v_t^*)=\arg\n\max_{\3n\3n\3n v_0,\cdots,v_t}\bigg\{\E[J_{t+1}(x_{t+1};v_0,v_1,\cdots,v_{t})|\mathcal{F}_0]+\sum_{j=0}^{t} \E[h_j(x_j,v_j)|\mathcal{F}_{0}]\bigg\}.
\end{align*}
Thus, we have
\begin{align*}
&~~~~(v_0^*,v_1^*,\cdots,v_t^*) \\
&=\arg\n\max_{\3n\3n\3n v_0,\cdots,v_t}\bigg\{\E[\E[J_{t+1}(x_{t+1};v_0,v_1,\cdots,v_{t})|\mathcal{F}_t]|\mathcal{F}_0]+\sum_{j=0}^{t} \E[h_j(x_j,v_j)|\mathcal{F}_{0}]\bigg\} \\
&=\arg\n\max_{\3n\3n\3n v_0,\cdots,v_t}\bigg\{\E[G_t^1(x_t;v_0,v_1,\cdots,v_{t})+G_t^2(x_t;v_0,v_1,\cdots,v_{t})|\mathcal{F}_0]+\sum_{j=0}^{t} \E[h_j(x_j,v_j)|\mathcal{F}_{0}]\bigg\} \\
&=\arg\n\max_{\3n\3n\3n v_0,\cdots,v_t}\bigg\{\E[G_t^1(x_t;v_0,v_1,\cdots,v_{t})|\mathcal{F}_0]+\sum_{j=0}^{t} \E[h_j(x_j,v_j)|\mathcal{F}_{0}]\bigg\} \\
&=\arg\n\max_{\3n\3n\3n v_0,\cdots,v_t}\Big\{\E[\cdots\E[\E[G_t^1(x_t;v_0,v_1,\cdots,v_{t})+h_t(x_t,v_t)|\mathcal{F}_{t-1}]+h_{t-1}(x_{t-1},v_{t-1})|\mathcal{F}_{t-2}]\cdots|\mathcal{F}_0]\\
&\quad\quad\quad\quad\quad\quad+h_0(x_0,v_0)\Big\},
\end{align*}
which implies
\begin{align*}
v_{t}^*=\arg\n\max_{\5n\5n v_t}\big\{G_t^1(x_t;v_0,v_1,\cdots,v_{t})+h_t(x_t,v_t)\big\}.
\end{align*}
Since $\E[G_t^2(x_t;v_0,v_1,\cdots,v_{t})|\mathcal{F}_0]=0$ holds for any admissible $(v_0,v_1,\cdots,v_{t})$, we have
\begin{align*}
J_0(x_0)&=\max_{v_0,\cdots,v_{t}}\bigg\{\E[J_{t+1}(x_{t+1};v_0,v_1,\cdots,v_{t})|\mathcal{F}_0]+\sum_{j=0}^{t} \E[h_j(x_j,v_j)|\mathcal{F}_{0}]\bigg\} \\
&=\max_{v_0,\cdots,v_{t}}\bigg\{\E[G_t^1(x_t;v_0,v_1,\cdots,v_{t})|\mathcal{F}_0]+\sum_{j=0}^{t} \E[h_j(x_j,v_j)|\mathcal{F}_{0}]\bigg\}.
\end{align*}
\endproof

\begin{remark}\label{rem:DP}
\sl
Please note that if $h_t(x_t,v_t)=h_t(x_t)$, i.e., $h_t$ is independent of control $v_t$, the conclusion of Lemma \ref{lem:DP} can be expressed as follows,
\begin{align*}
&v_{t}^*=\arg\n\max_{v_t}~G_t^1(x_t;v_0,v_1,\cdots,v_{t}), \\
&J_0(x_0)=\max_{v_0,\cdots,v_{t}}\bigg\{\E[G_t^1(x_t;v_0,v_1,\cdots,v_{t})|\mathcal{F}_0]+\sum_{j=0}^{t} \E[h_j(x_j)|\mathcal{F}_{0}]\bigg\}, \\
& \quad\quad\quad\quad\quad\quad\quad\quad\quad\quad\quad\quad\quad\quad\quad\quad\quad\quad\quad\quad t = 0,1,\cdots,T-1,
\end{align*}
i.e., $G_t^1(x_t;v_0,v_1,\cdots,v_{t}^*)+h_{t}(x_{t})$ can be regarded as the benefit-to-go function at time $t$.
\end{remark}

In this section, we reconsider the classical multi-period mean-variance model in Li
and Ng \cite{LiNg2000} and the multi-period mean-variance model with
intertemporal restrictions, $(MMV-MF)$, in Costa and Nabholz
\cite{CostaNabholz} under a mean-field formulation. Before presenting our main proposition, we define the following backwards recursions for $p_t$ and $q_t$,
\begin{align*}
\left\{\begin{array}{l}
p_t =  \alpha_t\rho_t+s_t^2(1-B_t)p_{t+1}, \\
p_T =  \alpha_T\rho_T,
\end{array}\right.\quad \left\{\begin{array}{l}
q_t =  \alpha_t\ell_t+s_tq_{t+1}, \\
q_T =  \alpha_T\ell_T,
\end{array}\right.
\end{align*}
for $t=T-1,T-2,\cdots,1$. We also set $\prod_{\emptyset}(\cdot)$ = $1$ and
$\sum_{\emptyset}(\cdot)$ = $0$ for the
convenience.

\begin{proposition}\label{prop_main}
\sl
The optimal strategy of problem $(MMV-MF)$ is given by
\begin{align}\label{eqn_opt_u_minus_exp}
\mathbf{u}_t^*-\mathbb{E}(\mathbf{u}_t^*)&=-s_t\big(x_t-\mathbb{E}(x_t)\big)\mathbb{E}^{-1}(\P_t\P_t')\mathbb{E}(\P_t), \\
\label{eqn_opt_exp_u}\mathbb{E}(\mathbf{u}_t^*)&=\frac{q_{t+1}}{2p_{t+1}}\frac{\mathbb{E}^{-1}(\P_t\P_t')\mathbb{E}(\P_t)}{1-B_t},
\end{align}
for $t=0,1,\cdots,T-1$, where the optimal expected wealth level is
\begin{align*}
\mathbb{E}(x_t)&=x_0\prod_{k=0}^{t-1}s_k+\sum_{j=0}^{t-1}\frac{q_{j+1}}{2p_{j+1}}\cdot\frac{B_j}{1-B_j}\cdot\prod_{\ell=j+1}^{t-1}s_\ell.
\end{align*}
\end{proposition}

\proof We first prove that, for information set
$\mathcal{F}_t=\sigma(\mathcal{F}_0\vee\sigma(\E(x_t),x_t-\E(x_t)))$,
we have the following expression,
\begin{align}
J_t\big(\E(x_t),x_t-\E(x_t)\big)=-p_t\big(x_{t}-\mathbb{E}(x_{t})\big)^2+q_t \mathbb{E}(x_{t})+\sum_{j=t}^{T-1}\frac{q^2_{j+1}}{4p_{j+1}}B_j, \label{eqn_opt_obj}
\end{align}
as the benefit-to-go function at time $t$.

When $t=T$, expression (\ref{eqn_opt_obj}) is obvious. Assume that
we have expression (\ref{eqn_opt_obj}) as the benefit-to-go function
at time $t+1$. We prove that expression (\ref{eqn_opt_obj}) still holds for the benefit-to-go function at time $t$.
For given information set $\mathcal{F}_t$, i.e.,
$(\E(x_t),x_t-\E(x_t))$, the recursive equation reads as
\begin{align*}
  & ~~ J_t\big(\E(x_t),x_t-\E(x_t)\big) \\
= &-\alpha_t\rho_t\big(x_{t}-\mathbb{E}(x_{t})\big)^2+\alpha_t\ell_t \mathbb{E}(x_{t})+\max_{(\E(\u_t),\u_t-\E(\u_t))}\E\big[J_{t+1}\big(\E(x_{t+1}),x_{t+1}-\E(x_{t+1})\big)\big|\mathcal{F}_t\big].
\end{align*}
Based on dynamics (\ref{eqn_expect_wealth}) and (\ref{eqn_wealth_minus_expect}), we deduce
\begin{align*}
&\E\big[J_{t+1}(\E(x_{t+1}),x_{t+1}-\E(x_{t+1}))\big|\mathcal{F}_t\big] \\
=&\E\big[-p_{t+1}\big(x_{t+1}-\mathbb{E}(x_{t+1})\big)^2 + q_{t+1} \mathbb{E}(x_{t+1})\big|\mathcal{F}_t\big]+\sum_{j=t+1}^{T-1}\frac{q^2_{j+1}}{4p_{j+1}}B_j \\
= & -p_{t+1}
\mathbb{E}\Big[s_{t}^2\big(x_t-\mathbb{E}(x_t)\big)^2 +\Big(\P_t'\big(\u_t-\mathbb{E}(\u_t)\big)\Big)^2+\Big(\mathbb{E}(\u_t')\big(\P_t-\mathbb{E}(\P_t)\big)\Big)^2\\
&+ 2s_t\big(x_t-\mathbb{E}(x_t)\big)\P_t'\big(\u_t-\mathbb{E}(\u_t)\big)+ 2s_t\big(x_t-\mathbb{E}(x_t)\big)\big(\P_t'-\mathbb{E}(\P_t')\big)\mathbb{E}(\u_t)\\
&+2\big(\u_t-\mathbb{E}(\u_t)\big)'\P_t\big(\P_t'-\mathbb{E}(\P_t')\big)\mathbb{E}(\u_t)\Big|\mathcal{F}_t\Big]+q_{t+1}\big[s_t\mathbb{E}(x_t)+\mathbb{E}(\P_t')\mathbb{E}(\u_t)\big]+\sum_{j=t+1}^{T-1}\frac{q^2_{j+1}}{4p_{j+1}}B_j.
\end{align*}
Since both $\u_t-\E(\u_t)$ and $\E(\u_t)$ are
$\mathcal{F}_t$-measurable and $\P_t$ is independent to
$\mathcal{F}_t$, we have
\begin{align*}
&\mathbb{E}\Big[\Big(\P_t'\big(\u_t-\mathbb{E}(\u_t)\big)\Big)^2\Big|\mathcal{F}_t\Big]=
\big(\u_t-\mathbb{E}(\u_t)\big)^\prime\E(\P_t\P_t')\big(\u_t-\mathbb{E}(\u_t)\big),\\
&\mathbb{E}\Big[\Big(\mathbb{E}(\u_t')\big(\P_t-\mathbb{E}(\P_t)\big)\Big)^2\Big|\mathcal{F}_t\Big]=
\mathbb{E}(\u_t')\big(\mathbb{E}(\P_t\P_t')-\mathbb{E}(\P_t)\mathbb{E}(\P_t')\big)\mathbb{E}(\u_t), \\
&\mathbb{E}\Big[2s_t\big(x_t-\mathbb{E}(x_t)\big)\P_t'\big(\u_t-\mathbb{E}(\u_t)\big)\mathbb{E}(\u_t)\Big|\mathcal{F}_t\Big]=2s_t\big(x_t-\mathbb{E}(x_t)\big)\E(\P_t')\big(\u_t-\mathbb{E}(\u_t)\big),\\
&\mathbb{E}\Big[2s_t\big(x_t-\mathbb{E}(x_t)\big)\big(\P_t'-\mathbb{E}(\P_t')\big)\mathbb{E}(\u_t)\Big|\mathcal{F}_t\Big]=0,\\
&\mathbb{E}\Big[2\big(\u_t-\mathbb{E}(\u_t)\big)'\P_t\big(\P_t'-\mathbb{E}(\P_t')\big)\mathbb{E}(\u_t)\Big|\mathcal{F}_t\Big]=2\big(\u_t-\mathbb{E}(\u_t)\big)'\big(\E(\P_t\P_t')-\E(\P_t)\mathbb{E}(\P_t')\big)\mathbb{E}(\u_t),
\end{align*}
which further implies,
\begin{align*}
&\E[J_{t+1}(\E(x_{t+1}),x_{t+1}-\E(x_{t+1}))|\mathcal{F}_t]\\
= & -p_{t+1}\Big[s_{t}^2\big(x_t-\mathbb{E}(x_t)\big)^2 +\big(\u_t-\mathbb{E}(\u_t)\big)^\prime\E(\P_t\P_t')\big(\u_t-\mathbb{E}(\u_t)\big) \\ &+2s_t\big(x_t-\mathbb{E}(x_t)\big)\E(\P_t^\prime)\big(\u_t-\mathbb{E}(\u_t)\big)\Big] -p_{t+1}\mathbb{E}(\u_t')\big(\mathbb{E}(\P_t\P_t')-\mathbb{E}(\P_t)\mathbb{E}(\P_t')\big)\mathbb{E}(\u_t) \\
&+q_{t+1}\mathbb{E}(\P_t')\mathbb{E}(\u_t)+s_tq_{t+1}\mathbb{E}(x_t)+\sum_{j=t+1}^{T-1}\frac{q^2_{j+1}}{4p_{j+1}}B_j\\
&+2\big(\u_t-\mathbb{E}(\u_t)\big)'\big(\E(\P_t\P_t')-\E(\P_t)\mathbb{E}(\P_t')\big)\mathbb{E}(\u_t)\\
=& G_t^1(\E(x_t),x_t-\E(x_t);\E(\u_t),\u_t-\E(\u_t))+G_t^2(\E(x_t),x_t-\E(x_t);\E(\u_t),\u_t-\E(\u_t)),
\end{align*}
where
\begin{align*}
  & G_t^1(\E(x_t),x_t-\E(x_t);\E(\u_t),\u_t-\E(\u_t)) \\
= & -p_{t+1}\Big[s_{t}^2\big(x_t-\mathbb{E}(x_t)\big)^2 +\big(\u_t-\mathbb{E}(\u_t)\big)^\prime\E(\P_t\P_t')\big(\u_t-\mathbb{E}(\u_t)\big) \\ &+2s_t\big(x_t-\mathbb{E}(x_t)\big)\E(\P_t^\prime)\big(\u_t-\mathbb{E}(\u_t)\big)\Big] -p_{t+1}\mathbb{E}(\u_t')\big(\mathbb{E}(\P_t\P_t')-\mathbb{E}(\P_t)\mathbb{E}(\P_t')\big)\mathbb{E}(\u_t) \\
&+q_{t+1}\mathbb{E}(\P_t')\mathbb{E}(\u_t)+s_tq_{t+1}\mathbb{E}(x_t)+\sum_{j=t+1}^{T-1}\frac{q^2_{j+1}}{4p_{j+1}}B_j\\
  &\5n G_t^2(\E(x_t),x_t-\E(x_t);\E(\u_t),\u_t-\E(\u_t))
= 2\big(\u_t-\mathbb{E}(\u_t)\big)'\big(\E(\P_t\P_t')-\E(\P_t)\mathbb{E}(\P_t')\big)\mathbb{E}(\u_t).
\end{align*}
Note that any admissible $(\E(\u_t),\u_t-\E(\u_t))$ satisfies $\E(\u_t-\E(\u_t))=0$, which implies
\begin{align*}
  &  \E\big[G_t^2\big(\E(x_t),x_t-\E(x_t);\E(\u_t),\u_t-\E(\u_t)\big)\big|\mathcal{F}_0\big] \\
= & 2\E\big[\big(\u_t-\mathbb{E}(\u_t)\big)'\big(\E(\P_t\P_t')-\E(\P_t)\mathbb{E}(\P_t')\big)\mathbb{E}(\u_t)\big|\mathcal{F}_0\big]
= 0.
\end{align*}
Using Lemma \ref{lem:DP} and Remark \ref{rem:DP}, we get
\begin{align*}
(\E(\u_t^*),\u_t^*-\E(\u_t^*))
=\arg\n\max G_t^1\big(\E(x_t),x_t-\E(x_t);\E(\u_t),\u_t-\E(\u_t)\big).
\end{align*}
By means of Lemma \ref{lem:EP}, we deduce
\begin{align*}
&G_t^1\big(\E(x_t),x_t-\E(x_t);\E(\u_t),\u_t-\E(\u_t)\big) \\
=&-p_{t+1}\Big\{s_{t}^2(1-B_t)\big(x_t-\mathbb{E}(x_t)\big)^2+ \Big[\big(\u_t-\mathbb{E}(\u_t)\big)+s_t\big(x_t-\mathbb{E}(x_t)\big)\mathbb{E}^{-1}(\P_t\P_t')\mathbb{E}(\P_t')\Big]'\\[2mm]
&\cdot \mathbb{E}(\P_t\P_t')\Big[\big(\u_t-\mathbb{E}(\u_t)\big)+s_t\big(x_t-\mathbb{E}(x_t)\big)\mathbb{E}^{-1}(\P_t\P_t')\mathbb{E}(\P_t')\Big]\Big\} \\
&-p_{t+1} \left[\mathbb{E}(\u_t)-\frac{q_{t+1}}{2p_{t+1}}\frac{\mathbb{E}^{-1}(\P_t\P_t')\mathbb{E}(\P_t)}{1-B_t}\right]'\big(\mathbb{E}(\P_t\P_t')-\mathbb{E}(\P_t)\mathbb{E}(\P_t')\big) \\
&\cdot\left[\mathbb{E}(\u_t)-\frac{q_{t+1}}{2p_{t+1}}\frac{\mathbb{E}^{-1}(\P_t\P_t')\mathbb{E}(\P_t)}{1-B_t}\right]+\frac{q^2_{t+1}}{4p_{t+1}}B_t+s_tq_{t+1}\mathbb{E}(x_t)+\sum_{j=t+1}^{T-1}\frac{q^2_{j+1}}{4p_{j+1}}B_j.
\end{align*}
Thus,
\begin{align*}
\mathbf{u}_t^*-\mathbb{E}(\mathbf{u}_t^*)&=-s_t\big(x_t-\mathbb{E}(x_t)\big)\mathbb{E}^{-1}(\P_t\P_t')\mathbb{E}(\P_t),\\
\mathbb{E}(\mathbf{u}_t^*)&=\frac{q_{t+1}}{2p_{t+1}}\frac{\mathbb{E}^{-1}(\P_t\P_t')\mathbb{E}(\P_t)}{1-B_t},
\end{align*}
where the linear constraint $\E(\u_t^*-\E(\u_t^*))=\0$ automatically holds.
Therefore, based on Remark \ref{rem:DP}, we have
\begin{align*}
&G_t^1\big(\E(x_t),x_t-\E(x_t);\E(\u_t^*),\u_t^*-\E(\u_t^*)\big)-\alpha_t\rho_t\big(x_{t}-\mathbb{E}(x_{t})\big)^2+\alpha_t\ell_t \mathbb{E}(x_{t})\\
=& -p_t\big(x_{t}-\mathbb{E}(x_{t})\big)^2+q_t \mathbb{E}(x_{t})+\sum_{j=t}^{T-1}\frac{q^2_{j+1}}{4p_{j+1}}B_j
\end{align*}
as the benefit-to-go function at time $t$.

Substituting the optimal expected portfolio strategy
(\ref{eqn_opt_exp_u}) into dynamics (\ref{eqn_expect_wealth}), we
further deduce the following recursive relationship of the optimal
expected wealth level,
\begin{align*}
\mathbb{E}(x_{t+1})&=s_t\E(x_t)+\frac{q_{t+1}}{2p_{t+1}}\cdot\frac{B_t}{1-B_t},
\end{align*}
which implies
\begin{align*}
\mathbb{E}(x_t) = x_0\prod_{k=0}^{t-1}s_k+\sum_{j=0}^{t-1}\frac{q_{j+1}}{2p_{j+1}}\cdot\frac{B_j}{1-B_j}\cdot\prod_{\ell=j+1}^{t-1}s_\ell.
\end{align*}
\endproof

The optimal strategy obtained in Proposition \ref{prop_main} covers the exiting results in the literature as its special cases.

Case 1: Let $I_\alpha=\{T\}$, $\alpha_T\ell_T=1$,
$\alpha_T\rho_T=\omega_T>0$. Then, we have
\begin{align*}
p_t=\omega_T\prod_{j=t}^{T-1}s_j^2(1-B_j),\quad q_t=\prod_{j=t}^{T-1}s_j,
\end{align*}
which further implies
\begin{align*}
\mathbb{E}(x_t)&=\prod_{k=0}^{t-1}s_kx_0+\frac{1}{2\omega_T}\prod_{k=t}^{T-1}s_k^{-1}\sum_{j=0}^{t-1}B_j\prod_{\ell=j}^{T-1} (1-B_\ell)^{-1}\\
&=\prod_{k=0}^{t-1}s_kx_0+\frac{1}{2\omega_T}\prod_{k=t}^{T-1}s_k^{-1}\frac{1-\prod_{k=0}^{t-1}(1-B_k)}{\prod_{k=0}^{T-1}(1-B_k)},\\
\mathbb{E}(x_T)&=\prod_{k=0}^{T-1}s_kx_0+\frac{1}{2\omega_T}\cdot\frac{1-\prod_{k=0}^{T-1}(1-B_k)}{\prod_{k=0}^{T-1}(1-B_k)}.
\end{align*}
Therefore, we have
\begin{align}
\nonumber \mathbf{u}_t^*=&-s_t\mathbb{E}^{-1}(\P_t\P_t')\mathbb{E}(\P_t')x_t+s_t\mathbb{E}^{-1}(\P_t\P_t')\mathbb{E}(\P_t')\mathbb{E}(x_t)+\mathbb{E}(u_t^*) \\
=&-s_t\mathbb{E}^{-1}(\P_t\P_t')\mathbb{E}(\P_t')x_t+\mathbb{E}^{-1}(\P_t\P_t')\mathbb{E}(\P_t')\Bigg[x_0\prod_{k=0}^{T-1}s_k+\frac{1}{2\omega_T\prod_{k=0}^{T-1}(1-B_k)}\Bigg]\prod_{k=t+1}^{T-1}s_k^{-1},\label{eqn_opt_u}
\end{align}
which is the optimal portfolio strategy obtained in Li and Ng \cite{LiNg2000}.

Substituting (\ref{eqn_opt_exp_u}) and (\ref{eqn_opt_u}) to dynamics (\ref{eqn_wealth_minus_expect}) yields
\begin{align*}
\mathbb{E}\big(x_{t+1}-\mathbb{E}(x_{t+1})\big)^2= s_t^2 (1-B_t) \mathbb{E}\big(x_t-\mathbb{E}(x_t)\big)^2+  \frac{1}{\prod_{k=t+1}^{T-1}s_k^2(1-B_k)}\cdot\frac{B_t}{4\omega_T^2\prod_{k=t}^{T-1}(1-B_j)},
\end{align*}
which further implies
\begin{align*}
\mathbb{E}\big(x_{T}-\mathbb{E}(x_{T})\big)^2 = & \ds \sum_{j=0}^{T-1} \prod_{k=j+1}^{T-1} s_k^2 (1-B_k) \frac{1}{\prod_{k=j+1}^{T-1}s_k^2(1-B_k)}\cdot\frac{B_j}{4\omega_T^2\prod_{k=j}^{T-1}(1-B_j)}\\
= & \ds \frac{1}{4\omega_T^2}\sum_{j=0}^{T-1} B_j\prod_{k=j}^{T-1} (1-B_k)^{-1}\\
= & \ds \frac{1-\prod_{k=0}^{T-1}(1-B_k)}{4\omega_T^2\prod_{k=0}^{T-1}(1-B_k)}.
\end{align*}
Thus, the efficient frontier is given by
\begin{align*}
\mbox{Var}(x_T)=\mathbb{E}\big(x_{T}-\mathbb{E}(x_{T})\big)^2=\frac{\prod_{k=0}^{T-1}(1-B_k)}{1-\prod_{k=0}^{T-1}(1-B_k)}\bigg(\mathbb{E}(x_{T})-x_0\prod_{k=0}^{T-1}s_k\bigg)^2~~ \mbox{for } \mathbb{E}(x_{T}) \geq x_0\prod_{k=0}^{T-1}s_k,
\end{align*}
which is the same as the efficient frontier established in Li and Ng \cite{LiNg2000}.

Case 2: Let $I_{\alpha}=\{\tau_1,\cdots,\tau_\alpha\}$ with $\tau_\alpha=T$. Then we have the optimal portfolio strategy as follows,
\begin{align}
\mathbf{u}_t^*&=-s_tx_t\mathbb{E}^{-1}(\P_t\P_t')\mathbb{E}(\P_t)+s_t\mathbb{E}(x_t)\mathbb{E}^{-1}(\P_t\P_t')\mathbb{E}(\P_t)+\frac{q_{t+1}}{2p_{t+1}}\frac{\mathbb{E}^{-1}(\P_t\P_t')\mathbb{E}(\P_t)}{1-B_t},\label{eqn_opt_u_IR}
\end{align}
where
\begin{align*}
\left\{\begin{array}{l}
p_t =  \alpha_t\rho_t+s_t^2(1-B_t)p_{t+1}, ~~t=\tau_i,\\
p_t =  s_t^2(1-B_t)p_{t+1}, ~~\tau_{i-1}< t< \tau_i,\\
p_T =  \alpha_T\rho_T,
\end{array}\right.~ \left\{\begin{array}{l}
q_t =  \alpha_t\ell_t+s_tq_{t+1}, ~~t=\tau_i,\\
q_t =  s_tq_{t+1}, ~~\tau_{i-1}<t<\tau_i,\\
q_T =  \alpha_T\ell_T,
\end{array}\right.
\end{align*}
and
\begin{align*}
\E(x_{t+1})=s_t\E(x_t)+\frac{q_{t+1}}{2p_{t+1}}\frac{B_t}{1-B_t},
\end{align*}
which is the same as the result developed in Costa and Nabholz
\cite{CostaNabholz}. Note that Costa and Nabholz originally
studied a market consisting of all risky assets in their
investigation. When we introduce a riskless
asset into the market, parameters of $\mathcal{G}_i$,
$\mathcal{S}_i$, $\mathcal{A}_i$ and $\mathcal{D}_i$ defined in
(22), (23), (28) and (29), respectively, in Costa and Nabholz
\cite{CostaNabholz} have been modified to
\begin{align*}
\mathcal{G}_i=-2p_{\tau_i},\quad \mathcal{S}_i=-q_{\tau_{i}},\quad
\mathcal{A}_i=\prod_{k=\tau_i}^{\tau_{i+1}-1}s_k,\quad
\mathcal{D}_i=\frac{1-\prod_{k=\tau_{i}}^{\tau_{i+1}-1}(1-B_k)}{\prod_{k=\tau_i}^{\tau_{i+1}-1}(1-B_k)}\cdot\frac{q_{\tau_{i+1}}}{2p_{\tau_{i+1}}}.
\end{align*}

\section{Generalized Mean-Variance Strategy with Risk Control Over Bankruptcy}
In this section, we reconsider the generalized mean-variance model with
risk control over bankruptcy in Zhu et al. \cite{Zhu:2004} under the mean-field framework, i.e., we consider problem
$(L-MF(\omega))$ first. For $t=T-1,T-2,\cdots,1$, we define $\bar{p}_t$, $\eta_t$ and $\xi_t$ as follows,
\begin{align*}
\left\{\begin{array}{l}
\bar{p}_t =  \omega_t+s_t^2(1-B_t)\bar{p}_{t+1}, \\
\bar{p}_T =  \omega_T,
\end{array}\right.\quad \left\{\begin{array}{l}
\eta_t =  \omega_ta_t+s_t^2\zeta_{t+1}\eta_{t+1}, \\
\eta_T =  0,
\end{array}\right.\quad \left\{\begin{array}{l}
\xi_t =  -\omega_ta_tb_t+s_t\zeta_{t+1}\xi_{t+1}, \\
\xi_T =  \frac{1}{2},
\end{array}\right.
\end{align*}
where Lagrangian multiplier $\omega_t \geq 0$ and
\begin{align*}
\zeta_{t+1}=\frac{\bar{p}_{t+1}(1-B_t)+2\eta_{t+1}B_t}{\bar{p}_{t+1}(1-B_t)+\eta_{t+1}B_t}\geq 0,
\end{align*}
due to $1>B_t>0$. Then, it is obvious that $\bar{p}_{t}>0$ and $\eta_t\geq 0$.

\begin{lemma}\label{lem:EPP}
\sl
Suppose that $\bar{p}_{t+1}>0$ and $\eta_{t+1}\geq 0$ hold. Then
\begin{align*}
\big[\bar{p}_{t+1}\mathbb{E}(\P_t\P_t')-(\bar{p}_{t+1}+\eta_{t+1})\mathbb{E}(\P_t)\mathbb{E}(\P_t')\big]^{-1}\mathbb{E}(\P_t)=\frac{\mathbb{E}^{-1}(\P_t\P_t')\mathbb{E}(\P_t)}{\bar{p}_{t+1}(1-B_t)+\eta_{t+1}B_t}.
\end{align*}
\end{lemma}

\proof
Applying Sherman-Morrison formula (Lemma \ref{lem_S_M}) yields
\begin{align*}
 & \big[\bar{p}_{t+1}\mathbb{E}(\P_t\P_t')-(\bar{p}_{t+1}+\eta_{t+1})\mathbb{E}(\P_t)\mathbb{E}(\P_t')\big]^{-1}\mathbb{E}(\P_t) \\
=& \left[\bar{p}_{t+1}^{-1}\mathbb{E}^{-1}(\P_t\P_t')+\frac{\bar{p}_{t+1}^{-1}\mathbb{E}^{-1}(\P_t\P_t')(\bar{p}_{t+1}+\eta_{t+1})\mathbb{E}(\P_t)\mathbb{E}(\P_t')\bar{p}_{t+1}^{-1}\mathbb{E}^{-1}(\P_t\P_t')}{1-\bar{p}_{t+1}^{-1}(\bar{p}_{t+1}+\eta_{t+1})\mathbb{E}(\P_t')\mathbb{E}^{-1}(\P_t\P_t')\mathbb{E}(\P_t)}\right]\mathbb{E}(\P_t) \\
=& \left[\bar{p}_{t+1}^{-1}\mathbb{E}^{-1}(\P_t\P_t')+\frac{\bar{p}_{t+1}^{-1}\mathbb{E}^{-1}(\P_t\P_t')(\bar{p}_{t+1}+\eta_{t+1})\mathbb{E}(\P_t)\mathbb{E}(\P_t')\bar{p}_{t+1}^{-1}\mathbb{E}^{-1}(\P_t\P_t')}{1-\bar{p}_{t+1}^{-1}(\bar{p}_{t+1}+\eta_{t+1})B_t}\right]\mathbb{E}(\P_t) \\
=& \frac{\mathbb{E}^{-1}(\P_t\P_t')\mathbb{E}(\P_t)}{\bar{p}_{t+1}(1-B_t)+\eta_{t+1}B_t}.
\end{align*}
\endproof

\begin{proposition}\label{prop_main_bankruptcy}
\sl
The optimal strategy of problem $(L-MF(\omega))$ is given by
\begin{align}\label{eqn_opt_u_minus_exp_bankruptcy}
\mathbf{u}_t^*-\mathbb{E}(\mathbf{u}_t^*)&=-s_t\big(x_t-\mathbb{E}(x_t)\big)\mathbb{E}^{-1}(\P_t\P_t')\mathbb{E}(\P_t),\\
\label{eqn_opt_exp_u_bankruptcy}\mathbb{E}(\mathbf{u}_t^*)&=\frac{\xi_{t+1}+\eta_{t+1}s_t\E(x_t)}{\bar{p}_{t+1}(1-B_t)+\eta_{t+1}B_t}\mathbb{E}^{-1}(\P_t\P_t')\mathbb{E}(\P_t),
\end{align}
where the optimal expected wealth level $\mathbb{E}(x_t)$ evolves
according to
\begin{align}
\mathbb{E}(x_t)=x_0\prod_{k=0}^{t-1}\zeta_{k+1}s_k+\sum_{j=0}^{t-1}\frac{\xi_{j+1}B_j}{\bar{p}_{j+1}(1-B_j)+\eta_{j+1}B_j}\cdot\prod_{\ell=j+1}^{t-1}\zeta_{\ell+1}s_\ell.
\label{eqn_opt_exp_wealth}
\end{align}
Moreover, the optimal objective function of $(L-MF(\omega))$ is
\begin{align}
H(\omega)=\eta_{1}\zeta_{1}s_0^2 x_0^2 +2\xi_{1}\zeta_{1}s_0 x_0+\sum_{j=0}^{T-1}\left[\frac{\xi_{j+1}^2B_j}{\bar{p}_{j+1}(1-B_j)+\eta_{j+1}B_j}+\omega_ja_jb_j^2\right],
\label{eqn_opt_L_obj}
\end{align}
with $\omega_0=a_0=b_0=0$.
\end{proposition}

\proof We first prove that for information set
$\mathcal{F}_t=\sigma(\mathcal{F}_0\vee\sigma(\E(x_t),x_t-\E(x_t)))$,
we have the following expression,
\begin{align}
\nonumber &J_t(\E(x_t),x_t-\E(x_t))\\
=&-\bar{p}_t\big(x_{t}-\mathbb{E}(x_{t})\big)^2+\eta_t(\E(x_t))^2+2\xi_t\E(x_t)+\sum_{j=t}^{T-1}\left[\frac{\xi_{j+1}^2B_j}{\bar{p}_{j+1}(1-B_j)+\eta_{j+1}B_j}+\omega_ja_jb_j^2\right],
\label{eqn_opt_obj_bankruptcy}
\end{align}
as the benefit-to-go function at time $t$.

When $t=T$, expression (\ref{eqn_opt_obj_bankruptcy}) is obvious.
Assume that expression (\ref{eqn_opt_obj_bankruptcy}) holds at
time $t+1$ as the benefit-to-go function. We show that expression (\ref{eqn_opt_obj_bankruptcy}) still holds for
the benefit-to-go function at time $t$. For given information set
$\mathcal{F}_t$, i.e., $(\E(x_t),x_t-\E(x_t))$, applying the
recursive equation yields
\begin{align*}
J_t(\E(x_t),x_t-\E(x_t))=&
-\omega_t\big(x_{t}-\mathbb{E}(x_{t})\big)^2
+\omega_ta_t(\E(x_t))^2-2\omega_ta_tb_t\E(x_t)+\omega_ta_tb_t^2\\
\nonumber&+\max_{(\E(\u_t),\u_t-\E(\u_t))}\E\left[J_{t+1}(\E(x_{t+1}),x_{t+1}-\E(x_{t+1}))\big|\mathcal{F}_t\right].
\end{align*}
Based on the dynamics in (\ref{eqn_expect_wealth}) and
(\ref{eqn_wealth_minus_expect}), we have
\begin{align*}
&\E\left[J_{t+1}(\E(x_{t+1}),x_{t+1}-\E(x_{t+1}))\big|\mathcal{F}_t\right]\\
=&\E\left[-\bar{p}_{t+1}\big(x_{t+1}-\mathbb{E}(x_{t+1})\big)^2+\eta_{t+1}(\E(x_{t+1}))^2+2\xi_{t+1}\E(x_{t+1})\big|\mathcal{F}_t\right]\\
&+\sum_{j=t+1}^{T-1}\left[\frac{\xi_{j+1}^2B_j}{\bar{p}_{j+1}(1-B_j)+\eta_{j+1}B_j}+\omega_ja_jb_j^2\right] \\
= & -\bar{p}_{t+1}
\mathbb{E}\Big[s_{t}^2\big(x_t-\mathbb{E}(x_t)\big)^2 +\Big(\P_t'\big(\u_t-\mathbb{E}(\u_t)\big)\Big)^2+\Big(\mathbb{E}(\u_t')\big(\P_t-\mathbb{E}(\P_t)\big)\Big)^2\\
&+ 2s_t\big(x_t-\mathbb{E}(x_t)\big)\P_t'\big(\u_t-\mathbb{E}(\u_t)\big)+ 2s_t\big(x_t-\mathbb{E}(x_t)\big)\big(\P_t'-\mathbb{E}(\P_t')\big)\mathbb{E}(\u_t)\\
& + 2\big(\u_t-\mathbb{E}(\u_t)\big)'\P_t\big(\P_t'-\mathbb{E}(\P_t')\big)\mathbb{E}(\u_t)\Big|\mathcal{F}_t\Big]+\eta_{t+1}\Big[s_t\mathbb{E}(x_t)+\mathbb{E}(\P_t')\mathbb{E}(\u_t)\Big]^2\\
& +2\xi_{t+1}\Big[s_t\mathbb{E}(x_t)+\mathbb{E}(\P_t')\mathbb{E}(\u_t)\Big]+\sum_{j=t+1}^{T-1}\left[\frac{\xi_{j+1}^2B_j}{\bar{p}_{j+1}(1-B_j)+\eta_{j+1}B_j}+\omega_ja_jb_j^2\right].
\end{align*}
Similar to the proof of Proposition \ref{prop_main}, we have
\begin{align*}
  &\E\left[J_{t+1}(\E(x_{t+1}),x_{t+1}-\E(x_{t+1}))\big|\mathcal{F}_t\right] \\
= & -\bar{p}_{t+1}\Big[s_{t}^2\big(x_t-\mathbb{E}(x_t)\big)^2 +\big(\u_t-\mathbb{E}(\u_t)\big)^\prime\E(\P_t\P_t')\big(\u_t-\mathbb{E}(\u_t)\big) \\
  &+2s_t\big(x_t-\mathbb{E}(x_t)\big)\E(\P_t^\prime)\big(\u_t-\mathbb{E}(\u_t)\big)\Big]-\mathbb{E}(\u_t')\Big[\bar{p}_{t+1}\mathbb{E}(\P_t\P_t')-(\bar{p}_{t+1}+\eta_{t+1})\mathbb{E}(\P_t)\mathbb{E}(\P_t')\Big]\mathbb{E}(\u_t) \\
  &+(2\xi_{t+1}+2\eta_{t+1}s_t\E(x_t))\mathbb{E}(\P_t')\mathbb{E}(\u_t)+\eta_{t+1}s_t^2(\E(x_t))^2+2\xi_{t+1}s_t\E(x_t)\\
  &+\sum_{j=t+1}^{T-1}\left[\frac{\xi_{j+1}^2B_j}{\bar{p}_{j+1}(1-B_j)+\eta_{j+1}B_j}+\omega_ja_jb_j^2\right]+2\big(\u_t-\mathbb{E}(\u_t)\big)'\big(\E(\P_t\P_t')-\E(\P_t)\mathbb{E}(\P_t')\big)\mathbb{E}(\u_t)\\
  =& G_t^1(\E(x_t),x_t-\E(x_t);\E(\u_t),\u_t-\E(\u_t))+ G_t^2(\E(x_t),x_t-\E(x_t);\E(\u_t),\u_t-\E(\u_t)),
\end{align*}
where
\begin{align*}
&G_t^1(\E(x_t),x_t-\E(x_t);\E(\u_t),\u_t-\E(\u_t))\\
=&-\bar{p}_{t+1}\Big[s_{t}^2\big(x_t-\mathbb{E}(x_t)\big)^2 +\big(\u_t-\mathbb{E}(\u_t)\big)^\prime\E(\P_t\P_t')\big(\u_t-\mathbb{E}(\u_t)\big) \\
  &+2s_t\big(x_t-\mathbb{E}(x_t)\big)\E(\P_t^\prime)\big(\u_t-\mathbb{E}(\u_t)\big)\Big]-\mathbb{E}(\u_t')\Big[\bar{p}_{t+1}\mathbb{E}(\P_t\P_t')-(\bar{p}_{t+1}+\eta_{t+1})\mathbb{E}(\P_t)\mathbb{E}(\P_t')\Big]\mathbb{E}(\u_t) \\
  &+(2\xi_{t+1}+2\eta_{t+1}s_t\E(x_t))\mathbb{E}(\P_t')\mathbb{E}(\u_t)+\eta_{t+1}s_t^2(\E(x_t))^2+2\xi_{t+1}s_t\E(x_t)\\
  &+\sum_{j=t+1}^{T-1}\left[\frac{\xi_{j+1}^2B_j}{\bar{p}_{j+1}(1-B_j)+\eta_{j+1}B_j}+\omega_ja_jb_j^2\right],\\
&\5n G_t^2(\E(x_t),x_t-\E(x_t);\E(\u_t),\u_t-\E(\u_t))=2\big(\u_t-\mathbb{E}(\u_t)\big)'\big(\E(\P_t\P_t')-\E(\P_t)\mathbb{E}(\P_t')\big)\mathbb{E}(\u_t).
\end{align*}
Note that any admissible $(\E(\u_t),\u_t-\E(\u_t))$ satisfies $\E(\u_t-\E(\u_t))=0$, which implies
\begin{align*}
\E\big[G_t^2\big(\E(x_t),x_t-\E(x_t);\E(\u_t),\u_t-\E(\u_t))\big|\mathcal{F}_0\big]=0.
\end{align*}
Using Lemma \ref{lem:DP} and corresponding to Remark \ref{rem:DP}, we get
\begin{align*}
(\E(\u_t^*),\u_t^*-\E(\u_t^*))
=\arg\n\max G_t^1\big(\E(x_t),x_t-\E(x_t);\E(\u_t),\u_t-\E(\u_t)\big).
\end{align*}
By means of Lemma \ref{lem:EP}, we deduce
\begin{align*}
  &  G_t^1(\E(x_t),x_t-\E(x_t);\E(\u_t),\u_t-\E(\u_t)) \\
= & - \bar{p}_{t+1}\mathbb{E}\Big\{ s_{t}^2(1-B_t)\big(x_t-\mathbb{E}(x_t)\big)^2+ \Big[\big(\u_t-\mathbb{E}(\u_t)\big)+s_t\big(x_t-\mathbb{E}(x_t)\big)\mathbb{E}^{-1}(\P_t\P_t')\mathbb{E}(\P_t')\Big]'\mathbb{E}(\P_t\P_t') \\
  & \cdot \Big[\big(\u_t-\mathbb{E}(\u_t)\big)+s_t\big(x_t-\mathbb{E}(x_t)\big)\mathbb{E}^{-1}(\P_t\P_t')\mathbb{E}(\P_t')\Big]\Big\} \\
  & - \left[\mathbb{E}(\u_t)-\frac{\xi_{t+1}+\eta_{t+1}s_t\E(x_t)}{\bar{p}_{t+1}(1-B_t)+\eta_{t+1}B_t}\mathbb{E}^{-1}(\P_t\P_t')\mathbb{E}(\P_t)\right]'\Big[\bar{p}_{t+1}\mathbb{E}(\P_t\P_t')-(\bar{p}_{t+1}+\eta_{t+1})\mathbb{E}(\P_t)\mathbb{E}(\P_t')\Big] \\
  & \cdot\left[\mathbb{E}(\u_t)-\frac{\xi_{t+1}+\eta_{t+1}s_t\E(x_t)}{\bar{p}_{t+1}(1-B_t)+\eta_{t+1}B_t}\mathbb{E}^{-1}(\P_t\P_t')\mathbb{E}(\P_t)\right]+\frac{(\xi_{t+1}+\eta_{t+1}s_t\E(x_t))^2}{\bar{p}_{t+1}(1-B_t)+\eta_{t+1}B_t}B_t\\
  & +\eta_{t+1}s_t^2(\E(x_t))^2+2\xi_{t+1}s_t\E(x_t)+\sum_{j=t+1}^{T-1}\left[\frac{\xi_{j+1}^2B_j}{\bar{p}_{j+1}(1-B_j)+\eta_{j+1}B_j}+\omega_ja_jb_j^2\right].
\end{align*}
Thus,
\begin{align*}
\mathbf{u}_t^*-\mathbb{E}(\mathbf{u}_t^*)&=-s_t\big(x_t-\mathbb{E}(x_t)\big)\mathbb{E}^{-1}(\P_t\P_t')\mathbb{E}(\P_t),\\
\mathbb{E}(\mathbf{u}_t^*)&=\frac{\xi_{t+1}+\eta_{t+1}s_t\E(x_t)}{\bar{p}_{t+1}(1-B_t)+\eta_{t+1}B_t}\mathbb{E}^{-1}(\P_t\P_t')\mathbb{E}(\P_t),
\end{align*}
which satisfies the linear constraint $\E(\u_t-\E(\u_t))=\0$.

Based on Remark \ref{rem:DP}, we can find
\begin{align*}
  & G_t^1(\E(x_t),x_t-\E(x_t);\E(\u_t^*),\u_t^*-\E(\u_t^*))\\
  &-\omega_t\big(x_{t}-\mathbb{E}(x_{t})\big)^2
+\omega_ta_t(\E(x_t))^2-2\omega_ta_tb_t\E(x_t)+\omega_ta_tb_t^2\\
= & -\bar{p}_t\big(x_{t}-\mathbb{E}(x_{t})\big)^2+\eta_t(\E(x_t))^2+2\xi_t\E(x_t)+\sum_{j=t}^{T-1}\left[\frac{\xi_{j+1}^2B_j}{\bar{p}_{j+1}(1-B_j)+\eta_{j+1}B_j}+\omega_ja_jb_j^2\right]
\end{align*}
as the benefit-to-go function at time $t$.

Substituting the optimal expected portfolio strategy (\ref{eqn_opt_exp_u_bankruptcy}) into dynamics (\ref{eqn_expect_wealth}) gives rise to
\begin{align*}
\mathbb{E}(x_{t+1})&=\zeta_{t+1}s_t\E(x_t)+\frac{\xi_{t+1}B_t}{\bar{p}_{t+1}(1-B_t)+\eta_{t+1}B_t},
\end{align*}
which implies
\begin{align*}
\mathbb{E}(x_t)=x_0\prod_{k=0}^{t-1}\zeta_{k+1}s_k+\sum_{j=0}^{t-1}\frac{\xi_{j+1}B_j}{\bar{p}_{j+1}(1-B_j)+\eta_{j+1}B_j}\cdot\prod_{\ell=j+1}^{t-1}\zeta_{\ell+1}s_\ell.
\end{align*}

Noting that $\omega_0=a_0=b_0=0$, then the expression of the optimal objective function of $(L-MF(\omega))$ is obvious.
\endproof

Substituting (\ref{eqn_opt_u_minus_exp_bankruptcy}) and (\ref{eqn_opt_exp_u_bankruptcy}) to dynamics (\ref{eqn_wealth_minus_expect}) yields
\begin{align*}
\mathbb{E}\big(x_{t+1}-\mathbb{E}(x_{t+1})\big)^2= s_t^2 (1-B_t) \mathbb{E}\big(x_t-\mathbb{E}(x_t)\big)^2+ \frac{\left(\xi_{t+1}+\eta_{t+1}s_t\E(x_t)\right)^2}{\left(\bar{p}_{t+1}(1-B_t)+\eta_{t+1}B_t\right)^2}(B_t-B_t^2),
\end{align*}
which leads to the following expression of the variance of the
optimal wealth level,
\begin{align*}
{\rm Var}(x_t)=\mathbb{E}\big(x_{t}-\mathbb{E}(x_{t})\big)^2 = \sum_{j=0}^{t-1}\frac{\left(\xi_{j+1}+\eta_{j+1}s_j\E(x_j)\right)^2}{\left(\bar{p}_{j+1}(1-B_j)+\eta_{j+1}B_j\right)^2}\cdot(B_j-B_j^2) \cdot\prod_{\ell=j+1}^{t-1}s_\ell^2(1-B_\ell).
\end{align*}

Zhu et al. \cite{Zhu:2004} analyzed the Lagrangian problem
$(L(\omega))$ via the embedding scheme. They, however, do not
succeed to obtain an analytical form of the optimal objective
value function $H(\omega)$. Thus, they proposed the prime-dual
algorithm to solve the following dual problem of $(GMV)$
numerically,
\begin{align*}
\min_{\omega\in\mathbb{R}^{T-1}_{+}} ~~H(\omega).
\end{align*}

In this paper, Proposition \ref{prop_main_bankruptcy} does not
only derive an analytical policy but also successfully reveal the
explicit form of $H(\omega)$. Thus, a simple steepest descent
algorithm can be directly applied to derive the optimal Lagrangian
multiplier vector $\omega^*$, due to the convexity of $H(\omega)$
(see \cite{Zhu:2004}). Then the optimal strategy of $(GMV)$ can be
presented by the portfolio strategy in Proposition
\ref{prop_main_bankruptcy} with $\omega=\omega^*$. Therefore, our
new mean-field formulation clearly, yet powerfully, offers a more
efficient and more accurate policy scheme again in this situation,
when compared to the existing literature.

\begin{exn} \sl
Consider an example of constructing a pension fund consisting of
S\&P 500 (SP), the index of Emerging Market (EM), Small Stock (MS)
of U.S market and a bank account. Based on the data provided in
Elton et al. \cite{Elton:2007}, the expected values, variances and
correlations of the annual return rates of these three indices are
given in Table \ref{Table1}.

\begin{table}[h]
  \centering
  \caption{Data for the asset allocation example}\label{Table1}
  \begin{tabular}{lccc}
    \hline
                    & SP       & EM      & MS \\
    \hline
    Expected Return & $14\%$   & $16\% $ & $17\%$ \\
    Variance        & $18.5\%$ & $30\% $ & $24\%$ \\
    \hline
    \multicolumn{4}{c}{Correlation}\\
    \hline
    SP              & $1$      & $0.64$     & $0.79$ \\
    EM              &          & $1$        & $0.75$ \\
    MS              &          &            & $1$ \\
    \hline
  \end{tabular}
\end{table}

We further assume that the annual risk free rate is $5\%$
($s_t=1.05$) and consider a five-period generalized mean-variance
model with risk control over bankruptcy, i.e., a $(GMV)$ problem.
Then, $\E(\P_t)$, $\mbox{\rm Cov}(\P_t)$ and $\E(\P_t\P_t')$ can
be computed as follows, for $t=0,1,\cdots,4$,
\begin{align*}
\E(\P_t)=\begin{bmatrix}
             0.09\\
             0.11\\
             0.12\\
         \end{bmatrix},
~
\mbox{\rm Cov}(\P_t)=\begin{bmatrix}
    0.0342  &  0.0355  &  0.0351\\
    0.0355  &  0.0900  &  0.0540\\
    0.0351  &  0.0540  &  0.0576\\
                                               \end{bmatrix},
~\E(\P_t\P_t')=\begin{bmatrix}
    0.0423  &  0.0454  &  0.0459 \\
    0.0454  &  0.1021  &  0.0672 \\
    0.0459  &  0.0672  &  0.0720 \\
                                               \end{bmatrix}.
\end{align*}

Assume that an investor has initial wealth $x_0=1$ and trade-off
parameter $\omega_5=1$. The disaster level and the acceptable
maximum probability of bankruptcy are chosen as $b_t=0$ and
$a_t=0.10$, respectively, for $t=1,2,3,4$.

To solve the dual problem of $(GMV)$ and get the optimal
Lagrangian multiplier vector $\omega^*$, we consider the following
unconstrained problem,
\begin{align*}
\min_{\omega\in\mathbb{R}^{4}} ~~H(\omega)-\mu \sum_{i=1}^{4}\log(\omega_i),
\end{align*}
where $\sum_{i=1}^{4}\log(\omega_i)$ is the barrier function used
to ensure $\omega$ $\in$ $\mathbb{R}_+^{4}$, $\mu$ is the barrier
parameter and $H(\omega)$ satisfies (\ref{eqn_opt_L_obj}).
Theoretically speaking, by setting $\mu\downarrow 0$, we can
derive the optimal Lagrangian multiplier vector. Using the
steepest descent algorithm, we get
\begin{align*}
\omega^*=[0.0014, 0.2658, 0.2543, 0.0014]^\prime.
\end{align*}
We further have
\begin{align*}
\begin{bmatrix}
\bar{p}_1\\
\bar{p}_2\\
\bar{p}_3\\
\bar{p}_4\\
\bar{p}_5\\
\end{bmatrix}=\begin{bmatrix}
0.9854\\
1.1364\\
1.0054\\
0.8673\\
1\\
\end{bmatrix},~~\begin{bmatrix}
\eta_1\\
\eta_2\\
\eta_3\\
\eta_4\\
\eta_5\\
\end{bmatrix}=\begin{bmatrix}
0.0615\\
0.0550\\
0.0256\\
0.0001\\
0\\
\end{bmatrix},~~\begin{bmatrix}
\xi_1\\
\xi_2\\
\xi_3\\
\xi_4\\
\xi_5\\
\end{bmatrix}=\begin{bmatrix}
0.6200\\
0.5828\\
0.5513\\
0.5250\\
0.5\\
\end{bmatrix},~~\begin{bmatrix}
\zeta_1\\
\zeta_2\\
\zeta_3\\
\zeta_4\\
\zeta_5\\
\end{bmatrix}=\begin{bmatrix}
1.0168\\
1.0130\\
1.0069\\
1.0000\\
1\\
\end{bmatrix}.
\end{align*}
The optimal expected wealth levels are then given by
\begin{align*}
\E(x_0)=1,~\E(x_1)=1.2366,~\E(x_2)=1.4537,~\E(x_3)=1.6856,~\E(x_4)=1.9353,~\E(x_5)=2.1687.
\end{align*}
Therefore, according to Proposition \ref{prop_main_bankruptcy},
the optimal strategy of $(GMV)$ is specified as follows,
\begin{align*}
&\u_0^*=(-1.05x_0+1.9197)\K, \\
&\u_1^*=(-1.05x_1+2.0218)\K, \\
&\u_2^*=(-1.05x_2+2.2688)\K, \\
&\u_3^*=(-1.05x_3+2.5409)\K, \\
&\u_4^*=(-1.05x_4+2.6687)\K,
\end{align*}
where
\begin{align*}
\K=\mathbb{E}^{-1}(\P_t\P_t')\mathbb{E}(\P_t)=\begin{bmatrix}
~~1.0580 \\
-0.1207 \\
~~1.1052 \\
   \end{bmatrix}.
\end{align*}
Finally, the variances of the optimal wealth levels are given as
\begin{align*}
{\rm Var}(x_1)= 0.1275,~{\rm Var}(x_2)= 0.1986,~{\rm Var}(x_3)= 0.2648,~{\rm Var}(x_4)=0.3295,~{\rm Var}(x_5)=0.3536.
\end{align*}

We can further get the efficient frontier of $(GMV)$ by adjusting
the trade-off parameter $\omega_5$ from $0$ to $+\infty$, which is
represented by the dash dot line in Figure
\ref{fig-two-frontiers}. In the figure, the solid curve above is
the efficient frontier of the classical five-period mean-variance
model, which is plotted for a comparison purpose.
\begin{figure}[h]
\centering
\includegraphics[height=8cm]{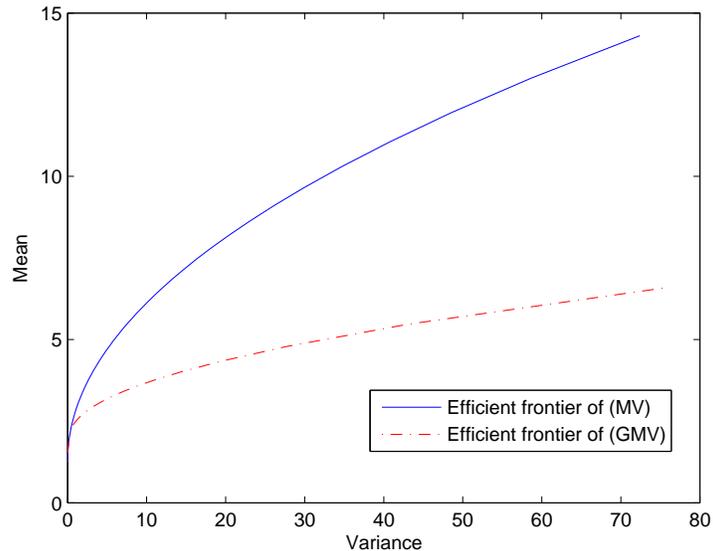}
\caption{Efficient frontiers of $(MV)$ and $(GMV)$} \label{fig-two-frontiers}
\end{figure}
\end{exn}

\section{Conclusions}

The nonseparable multi-period mean-variance and related problems
have been solved in the literature via embedding scheme,
Lagrangian formulation or mean-variance hedging problem. However,
we may not be able to derive optimal value functions of these
transformed problems analytically, especially, when some
constraints are attached to the problem setting. Hence, we often
need to invoke some numerical algorithms to compute the
corresponding best auxiliary parameter or Lagrangian parameter. In
this paper, we adopt the mean-filed formulation, as a more efficient means, to directly tackle
the nonseparable multi-period mean-variance portfolio selection
model, multi-period mean-variance model with intertemporal
restrictions, and generalized mean-variance model with risk
control over bankruptcy. Under this newly proposed framework of mean-field
formulations, we are capable of deriving analytical solutions for
all these problems, thus improving the solution quality and
facilitating the solution process.



\begin{thebibliography}{99}
\bibitem{Andersson-Djehiche}
D. Andersson, B. Djehiche,
A maximum principle for SDEs of mean-field type, \textit{Applied Mathematics and Optimization}, 63 (2011), 341-356.

\bibitem{Borkar-Kumar}
V.S. Borkar, K.S. Kumar,
McKean-Vlasov limit in portfolio optimization, \textit{Stochastic Processes and Their Applications}, 28 (2010), 884-906.

\bibitem{Buckdahn-Djehiche-Li}
R. Buckdahn, B. Djehiche, J. Li,
A general stochastic maximum principle for SDEs of mean-field type, \textit{Applied Mathematics and Optimization}, 64 (2011), 197-216.

\bibitem{Buckdahn-Li-Peng}
R. Buckdahn, J. Li, S. Peng,
Mean-field backward stochastic differential equations and related partial differential equations, \textit{Stochastic Processes and their Applications}, 119 (2009), 3133-3154.

\bibitem{Celikyurt-Ozekici}
U. \c{C}elikyurt, S. \"{O}zekici,
Multi-period portfolio optimization models in stochastic markets using the mean-variance approach, \textit{European Journal of Operational Research}, 179 (2007), 186-202.

\bibitem{CernyKallsen2009}
A. \v{C}ern\'{y}, J. Kellsen,
Hedging by sequential regressions revisted, \textit{Mathematical Finance}, 19 (2009), 591-617.

\bibitem{Chan}
T. Chan,
Dynamics of the McKean-Vlasov equation, \textit{Annals of Probability}, 22 (1994), 431-441.

\bibitem{ChenYang}
P. Chen, H.L. Yang,
Markowitz's mean-variance asset-liability management with regime switching: A multi-period model, \textit{Applied Mathematical Finance}, 18 (2011), 29-50.

\bibitem{ChiuLi}
M.C. Chiu, D. Li,
Asset and liability management under a continuous-time mean-variance optimization framework, \textit{Insurance: Mathematics and Economics}, 39 (2006), 330-355.

\bibitem{CostaNabholz}
O.L.V. Costa, R.B. Nabholz,
Multi-period mean-variance optimization with intertemporal restrictions, \textit{Journal of Optimization Theory and Applications}, 134 (2007), 257-274.

\bibitem{Crisan-Xiong}
D. Crisan, J. Xiong,
Approximate McKean-Vlasov representations for a class of SPDEs, \textit{Stochastics}, 82 (2010), 53-68.

\bibitem{Dawson}
D.A. Dawson,
Critical dynamics and fluctuations for a mean-field model of cooperative behavior, \textit{Journal of Statistical Physics}, 31 (1983), 29-85.

\bibitem{Elton:2007}
E.J. Elton,, M.J. Gruber, S.J. Brown, and W.N. Goetzmann,
\textit{Modern Portfolio Thoery and Investment Analysis}, John Wiley \& Sons, (2007).

\bibitem{Kac}
M. Kac,
Foundations of kinetic theory, \textit{Proceedings of the Third Berkeley Symposium on Mathematical Statistics and Probability}, 3 (1956), 171-197.

\bibitem{Leippold}
M. Leippold, F. Trojani, and P. Vanini,
A geometric approach to multi-period mean-variance optimization of assets and liabilities, \textit{Journal of Economic Dynamics and Control}, 28 (2004), 1079-1113.

\bibitem{LiNg2000}
D. Li, W.L. Ng,
Optimal dynamic portfolio selection: Multi-period mean-variance formulation, \textit{Mathematical Finance}, 10 (2000), 387-406.

\bibitem{LZL}
X. Li, X.Y. Zhou, A.E.B. Lim,
Dynamic mean-variance portfolio selection with no-shorting constraints, \textit{SIAM Journal on Control and Optimization}, 40 (2002), 1540-1555.

\bibitem{Markowitz}
H.M. Markowitz,
Portfolio selection, \textit{Journal of Finance}, 7 (1952), 77-91.

\bibitem{McKean}
H.P. McKean,
A class of Markov processes associated with nonlinear parabolic equations, \textit{Proceedings of the National Academy of Sciences of the United States of America}, 56 (1966), 1907-1911.


\bibitem{Merton}
R.C. Merton,
An analytic derivation of the efficient portfolio frontier, \textit{Journal of Financial and Quantitative Analysis}, 7 (1972), 1851-1872.

\bibitem{Meyer-Brandis-Oksendal-Zhou}
T. Meyer-Brandis, B. Oksendal, X. Y. Zhou,
A mean-field stochastic maximum principle via Malliavin calculus,
\textit{A special issue for Mark Davis' Festschrift, to appear in Stochastics}, (2011).

\bibitem{Nourian-Caines-Malhame-Huang}
M. Nourian, P.E. Caines, R.P. Malham\'e, M. Huang,
Nash, social and centralized solutions to consensus problems via mean field control theory,
\textit{to appear IEEE Transaction on Automatic Control}, (2012).

\bibitem{Schweizer1996}
M. Schweizer,
Approximation pricing and the variance-optimal martingale measure, \textit{Annals of Probability}, 24 (1996), 206-236.

\bibitem{SunWang}
W.G. Sun, C.F. Wang,
The mean-variance investment problem in a constrained financial market, \textit{Journal of Mathematical Economics}, 42 (2006), 885-895.

\bibitem{XiaYan2006}
J.M. Xia, J.A. Yan,
Markowitz's portfolio optimization in an incomplete market, \textit{Mathematical Finance}, 16 (2006), 203-216.

\bibitem{Yong2011}
J.M. Yong,
A linear-quadratic optimal control problem for mean-field stochastic differential equations, \textit{Working paper}, arXiv:1110.1564, (2012).

\bibitem{ZhouLi2000}
X.Y. Zhou, D. Li,
Continuous-time mean-variance portfolio selection: A stochastic LQ framework, \textit{Applied Mathematics and Optimization}, 42 (2000), 19-33.

\bibitem{Zhu:2004}
S.S. Zhu, D. Li, S.Y. Wang,
Risk control over bankruptcy in dynamic portfolio selection: A generalized mean-variance formulation, \textit{IEEE Transactions on Automatic Control}, 49 (2004), 447-457.

\end{thebibliography}
\end{document}